\documentclass[a4paper]{jpconf}

\pdfoutput=1

\usepackage{hyperref}
\usepackage{graphicx}
\usepackage{xspace}

\usepackage[T1]{fontenc}
\usepackage{lmodern}
\usepackage{microtype}

\hypersetup{pdfauthor={Johann Felix von Soden-Fraunhofen},pdftitle={GoSam 2.0: a tool for automated one-loop calculations},hidelinks}

\usepackage{cite}

\begin{document}
\title{GoSam 2.0: a tool for automated one-loop calculations}

\author{Johann Felix von Soden-Fraunhofen}
%
\address{Max Planck Institute for Physics, F\"ohringer Ring 6, 80805 Munich, Germany}
\ead{\href{mailto:jfsoden@mpp.mpg.de}{jfsoden@mpp.mpg.de}}

\newcommand{\GoSam}{\textsc{GoSam}\xspace}
\newcommand{\GoSamTwo}{\textsc{GoSam~2.0}\xspace}
\newcommand{\prgname}[1]{\textsc{#1}\xspace}

\begin{abstract}
We present the version 2.0 of the program package \GoSam, which
is a public program package to compute one-loop QCD and/or electroweak corrections to multi-particle processes within and beyond the Standard Model. The extended version of the Binoth-Les-Houches-Accord interface to Monte Carlo programs is also implemented. This allows a large flexibility regarding the combination of the code with various Monte Carlo programs to produce fully differential NLO results, including the possibility of parton showering and hadronisation. We illustrate the wide range of applicability of the code by showing phenomenological results for multi-particle processes at NLO, both within and beyond the Standard Model.
\end{abstract}
\section{Introduction}
In the last years, there has been a remarkable progress in the development of 
automated NLO tools for multi-particle processes.
This leads to the possibility of having NLO corrections matched to
parton-showers as the new standard at the LHC.

Here we want to summarize the new features of the \GoSamTwo release \cite{Cullen:2014yla},
which is able to automatically compute one-loop QCD and/or electroweak corrections to user-defined processes, 
and show the results of two applications. 

The developments since the former release of \GoSam described in \cite{Reiter:2009kb, Cullen:2011ac} has allowed
various phenomenological studies within and beyond the Standard model \cite{Greiner:2012im,vanDeurzen:2013rv,Gehrmann:2013aga,Luisoni:2013cuh,Hoeche:2013mua,Cullen:2013saa,vanDeurzen:2013xla,Gehrmann:2013bga, Dolan:2013rja,Heinrich:2013qaa, Cullen:2012eh, Greiner:2013gca}. The new features
are now publicly available in the \GoSamTwo release.

\section{GoSam~2.0 workflow}
\begin{figure}[htpb]
	\centering
	\includegraphics[width=0.4\linewidth]{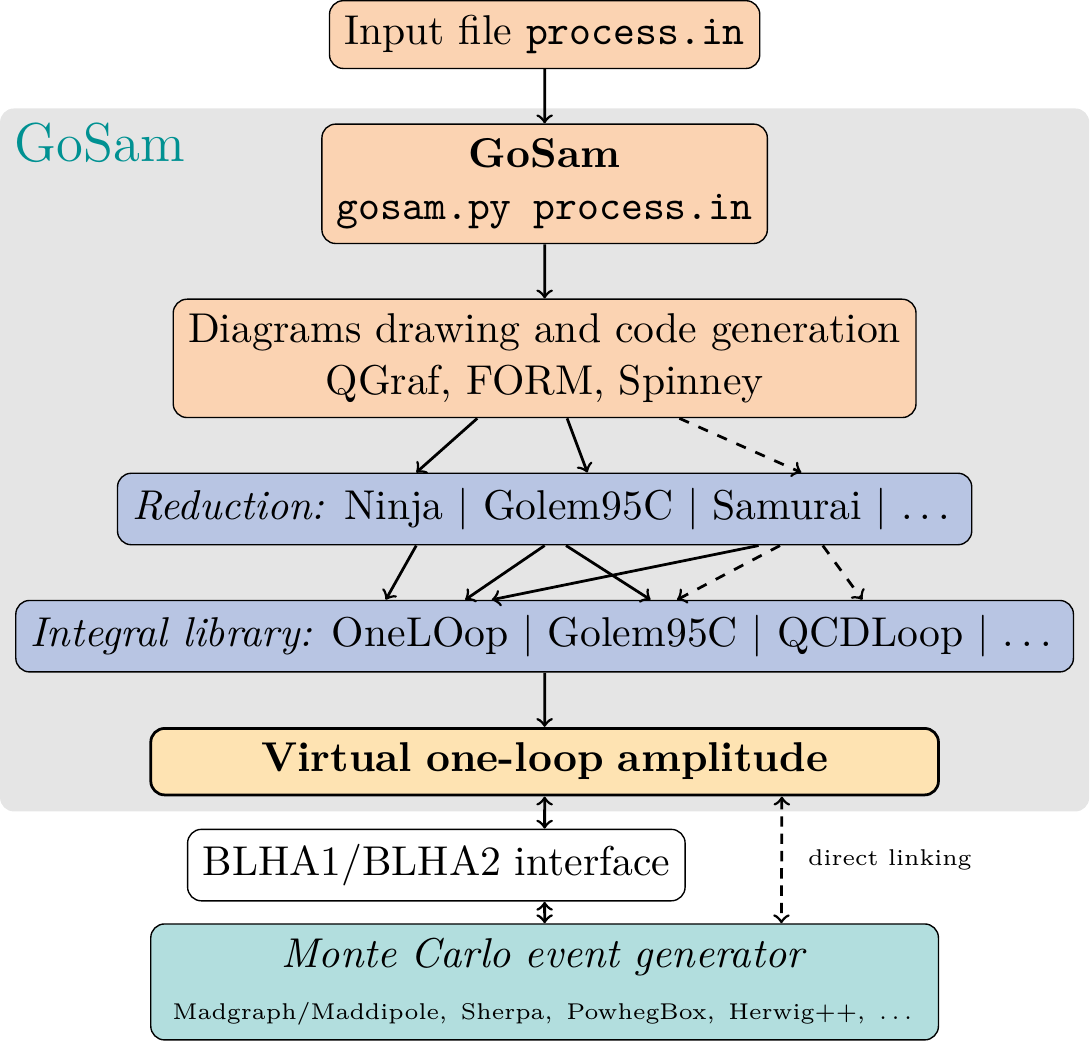}
	\caption{Basic workflow of \GoSam}
	\label{fig:GoSamWorkflow}
\end{figure}

\GoSam can be used standalone to generate code
for one-loop amplitudes\footnote{The generation of colour/spin correlated tree level amplitudes is also possible.}
or in combination with recent Monte Carlo programs via the Binoth-Les-Houches-Accord (BLHA) described in Sec.~\ref{ssec:blha2}.

In the standalone mode, which is shown schematically in Fig.~\ref{fig:GoSamWorkflow}, the user provides the needed information to \GoSam via an input card.
\GoSam calls \prgname{QGraf} \cite{Nogueira:1991ex} to generate all Feynman diagrams and the related algebraic expressions.
These are further processed by \prgname{FORM} \cite{Vermaseren:2000nd,Kuipers:2012rf}.
The spinor algebra is handled by the \prgname{Spinney} library\cite{Cullen:2010jv}.
The code is then transformed to compilable expressions using the new \prgname{FORM~4} features\cite{Kuipers:2012rf} (see Sec.~\ref{sec:newFeatures}).

At runtime, the generated code can choose between different integrand-level reduction methods \cite{Ossola:2006us,Giele:2008ve}, implemented in \prgname{Ninja} \cite{Mastrolia:2012bu,vanDeurzen:2013saa,Peraro:2014cba} and \prgname{Samurai} \cite{Mastrolia:2010nb,vanDeurzen:2013pja}, and improved tensor reduction methods implemented in \prgname{Golem95} \cite{Binoth:2005ff,Binoth:2008uq,Cullen:2011kv,Guillet:2013msa}.

\section{New features of GoSam~2.0}
\label{sec:newFeatures}
In this section we present some highlights of the version 2.0 of \GoSam.

\paragraph{Higher rank integrals}
While in calculations in Feynman gauge within in the Standard Model, the tensor rank is always smaller or equal
to the number of propagators in a loop integral, this is not necessarily the case in other gauges, effective theories or theories with
spin-2 particles. For the \GoSamTwo release, all of the supported reduction libraries (\prgname{Ninja}, \prgname{Samurai}, \prgname{Golem95}) have been extended
to the case where the rank of the tensor integrals exceeds the number of the propagators \cite{Mastrolia:2012bu,vanDeurzen:2013pja,Guillet:2013msa}.
\paragraph{New reduction method}
In \GoSamTwo, the new default reduction library is \prgname{Ninja} \cite{Mastrolia:2012bu,vanDeurzen:2013saa,Peraro:2014cba}.
\prgname{Ninja} is a new developed integrand-reduction method and exploits the fact that the integrands are provided by \GoSam
in analytic expressions. It uses the Laurent expansion of the integrands performed by semi-numerical polynomial divisions.
\paragraph{Improved code generation} The generation of optimized code is now
mainly handled by the new \prgname{FORM 4.0} optimization features \cite{Kuipers:2012rf,Kuipers:2013pba},
which enables faster and more compact code.
\paragraph{Grouping and summing of diagrams with common substructure}
\GoSamTwo has a new option \verb!diagsum! which automatically groups or combines diagrams which share a common loop-structure and/or which can be
distinguished only by external tree parts.
This reduces not only the code generation time, but also the number of calls to the reduction method and, therefore,
increases the overall performance. This extends the option of grouping diagrams already present in \GoSam 1.0.
\paragraph{Numerical polarisation vectors}
By default, \GoSamTwo creates code with generic polarisation 
vectors for massless gauge bosons and evaluates it for the
various helicity configurations numerically. This saves the
generation of separate code for each helicity configuration
and therefore reduces the code generation time and the 
disk and memory footprint.
\paragraph{Improved tensorial reconstruction}
The new \GoSam release enables by default the \verb!derive! extension, which
allows to isolate the tensor coefficients entering tensorial reconstruction
\cite{Heinrich:2010ax} in a more efficient way.  It is based on the fact that
the numerator can be written as a (tensorial) Taylor series in the loop momenta
$q^{\mu}$ around $q=0$.
\paragraph{Electroweak scheme choices}
\GoSamTwo supports various schemes defining which of the electroweak parameters are
input parameters and which ones are derived parameters.  From a certain set of
input parameters defining an electroweak scheme, GoSam will automatically determine the
derived parameters. If not otherwise specified, \GoSam uses the set ($M_W$,
$M_Z$, $\alpha$) by default.

The settings can be provided to \GoSam using the \verb!model.options! keyword, which also
needs to contain \verb!ewchoice! if the choice should be changeable at runtime. 
If \GoSam is accessed via the Binoth Les Houches interface (cf. Sec.~\ref{ssec:blha2}), the electroweak scheme
is determined automatically according to the \verb!OLP_SetParameter! calls.
\subsection{BLHA2 Interface}
\label{ssec:blha2}
\GoSamTwo supports the Binoth Les Houches Accord 2 (BLHA2), an updated standard interface between Monte Carlo event generators and one-loop programs
\cite{Alioli:2013nda}. The first version of the BLHA standard \cite{Binoth:2010xt} is still supported.

\begin{figure}[htpb]
	\centering
	\includegraphics[width=0.7\linewidth]{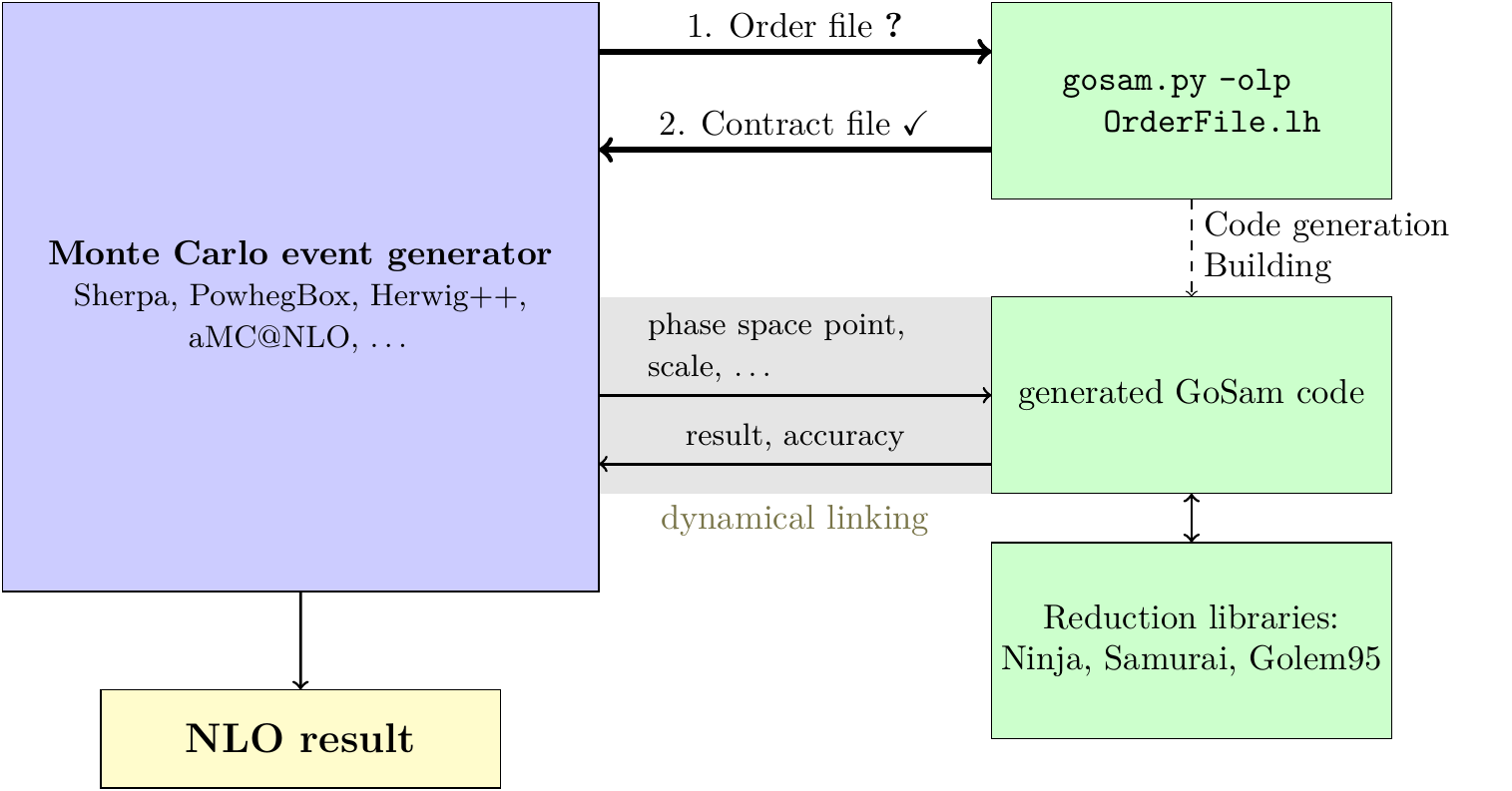}
	\caption{Workflow of the BLHA2 interface with \GoSam}
	\label{fig:BLHA2Workflow}
\end{figure}

The workflow of this interface is described in Fig.~\ref{fig:BLHA2Workflow}. It can be split into two phases. The first phase is the ``order/contract phase'', where 
the Monte Carlo programs ``asks'' the one-loop-provider (e.g. \GoSam) for the needed processes in an \textit{order file}. It can then ``reply'' with an
\textit{contract file} and generate the code.
As soon as the contract is ``signed'', i.\,e. all the items placed in the order
file can be provided, the second phase can start. The Monte Carlo program is
able to dynamically link the generated code for the virtual amplitude and can
evaluate the amplitude at phase space points, thus performing the integration
and event generation steps.

The updated interface allows to specify processes with different multiplicities and settings
in one order file. \GoSam can also provide colour- and spin-correlated LO matrix elements.
Both features allows the Monte Carlo program to built subtraction
terms of NLO real radiation in addition to the virtual part.

The accuracy can be assessed by various tests: 
The first test compares the infrared pole coefficients to those obtained from the general structure of infrared singularities.
The second, more time-consuming test is a rotation test, where the scattering amplitude is re-evaluated after an azimuthal rotation around the beam axis,
which should keep the amplitude invariant.
The thresholds triggering the various tests can be adjusted by the user via the input card.

These tests are also used to trigger the rescue system (switching to \prgname{Golem95} as reduction library) in the case 
of unstable results and are, if necessary, repeated to obtain the estimated accuracy of the final (rescued) result.

This procedure has been shown to be a good compromise between speed and quality of the assessed accuracy.

\section{Installation}
\GoSamTwo is distributed with a new install script, which installs the code automatically.
It also helps to update \GoSam and its components.

\noindent The install script can be downloaded by calling

\verb!wget http://gosam.hepforge.org/gosam-installer/gosam_installer.py!\\
The next step is to run the installer with the following commands

 \verb!chmod +x gosam_installer.py!

 \verb!./gosam_installer.py --prefix=path/to/install! \\
or alternatively:

  \verb!python gosam_installer.py --prefix=path/to/install!

\noindent If the \verb!--prefix! option is left out, the script installs \GoSam into a subdirectory \verb!./local! of the current working directory.
Upon installation, the installer asks the user if it should use existing
programs on the system (\prgname{FORM 4}, \prgname{QGraf}, reduction libraries) or should install new versions of them which are tested with \GoSam.
For a default installation, the questions can be safely answered by pressing the \verb!ENTER! key. 
After this, the installer downloads, builds and installs all components. Finally, a script
\verb![prefix]/bin/gosam_setup_env.sh! is created which sets up (temporarily) all environment variables
if sourced in a shell.

It is suggested to keep the \verb'installer-log.in', which may be needed to update
or uninstall \GoSam later. The update process can be initiated by calling the install script again without any arguments
in the same folder where this file is located.

\section{Using GoSam}
By calling \verb'gosam.py --template process.in', the user is able to generate an input file
where all possible settings are explained and default settings are shown. After filling all settings,
\verb'gosam.py process.in' generates a subdirectory specified by the \verb!process_path!
setting in the input card. The remaining steps for code generation
and building can be performed by calling \verb!make source! and \verb!make compile! in this
process directory. In the \verb!doc/! subfolder, \verb!make doc! generates documentation containing process specific
informations as well as Feynman diagrams. The \verb!matrix/! subdirectory contains a file \verb!test.f90!
which contains an example how to interface the compiled code. It can be built by calling \verb!make test.exe!
and outputs the NLO result for one random phase space point by default.

An example input-file for QCD corrections to $e^+e^-\,\to\,t\bar t$ is shown in Fig.~\ref{fig:eett}.

\begin{figure}[htbp]
\centering
\fbox{\parbox{0.3\textwidth}{\ttfamily\scriptsize\selectfont
process\_name=eett\\
process\_path=eett\_code\\
in=e+,e-\\
out=t,t~\\
order=QCD,0,2 \\
one=gs,e\\
zero=me,mU,mD,mS,mC,mB}}
\caption{Example of an input card for QCD corrections to $e^+e^-\,\to\,t\bar t$}
\label{fig:eett}
\end{figure}

\section{Applications}
\subsection{First example of \GoSam plus \prgname{Herwig++/Matchbox}} 

As a first example demonstrating the features of the BLHA2 interface implemented in 
\GoSamTwo and an upcoming version of \prgname{Herwig++/Matchbox} \cite{Bellm:2013lba,Platzer:2011bc, Bahr:2008pv},
the NLO QCD corrections to the process $pp\,\to\, Z/\gamma^* + \mbox{jet} \,\to e^+e^- + \mbox{jet}$ were calculated.

\begin{figure}[htpb]
	\centering
	\includegraphics[width=0.5\textwidth]{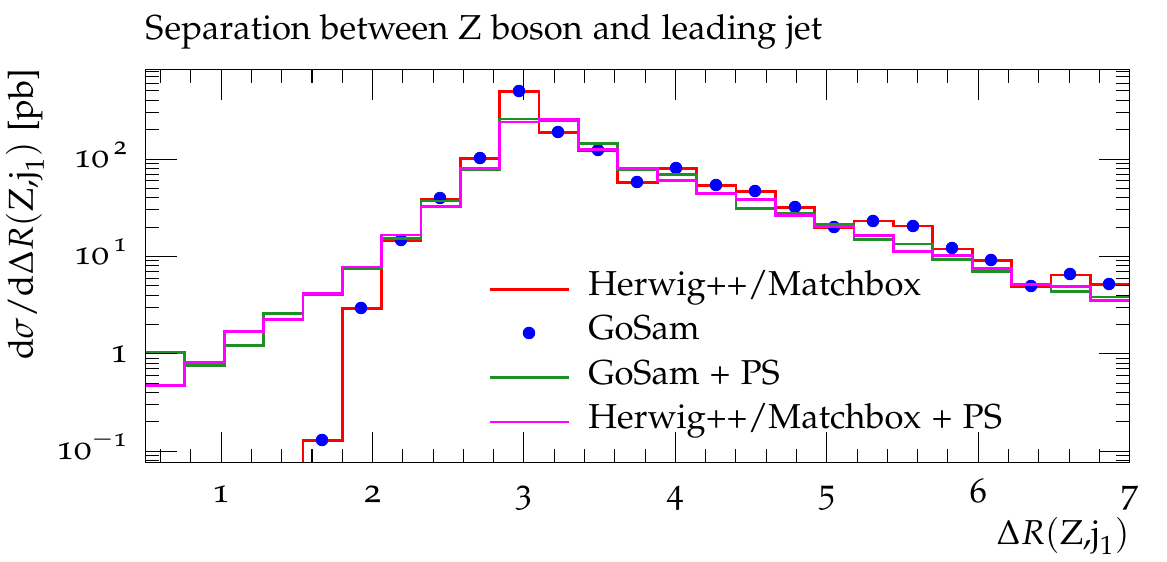}
	\caption{Z+jet production: R separation between the Z-boson and the leading jet. 
	 The blue dots and the green line show the result of \prgname{Herwig++/Matchbox} using matrix elements generated by \GoSam without and with 
	 parton shower; the red and pink lines show the corresponding results with built-in Herwig/Matchbox matrix elements.
	 }
	\label{fig:Zjet}
\end{figure}

Besides the virtual matrix elements, \GoSam provides various tree-level matrix elements (normal Born matrix elements and
spin- and colour-correlated Born matrix elements) via the BLHA2 interface, which supports now order files with subprocesses of different
amplitude types and multiplicities.
This allows Herwig++ to calculate the full NLO real radiation part. 

As this specific process is also directly built-in in \prgname{Herwig++/Matchbox}, it was easily possible to validate the results.
Fig.~\ref{fig:Zjet} shows the matching results. Both ways are largely automated, i.e. only small changes in the \prgname{Herwig++} input card (and optionally 
in an additional \GoSam setting card) are needed.

Furthermore, \prgname{Herwig++} can match a parton shower to the fixed-order calculation,
which gives access to new kinematical regions. The result is also plotted in Fig.~\ref{fig:Zjet}.

Full details of this calculation and information about the implementations can be found in \cite{Bellm:2014x1_jfs}.

\subsection{Diphoton plus jet production through graviton exchange}

\begin{figure}[htpb]
	\centering
	\includegraphics[width=0.5\textwidth]{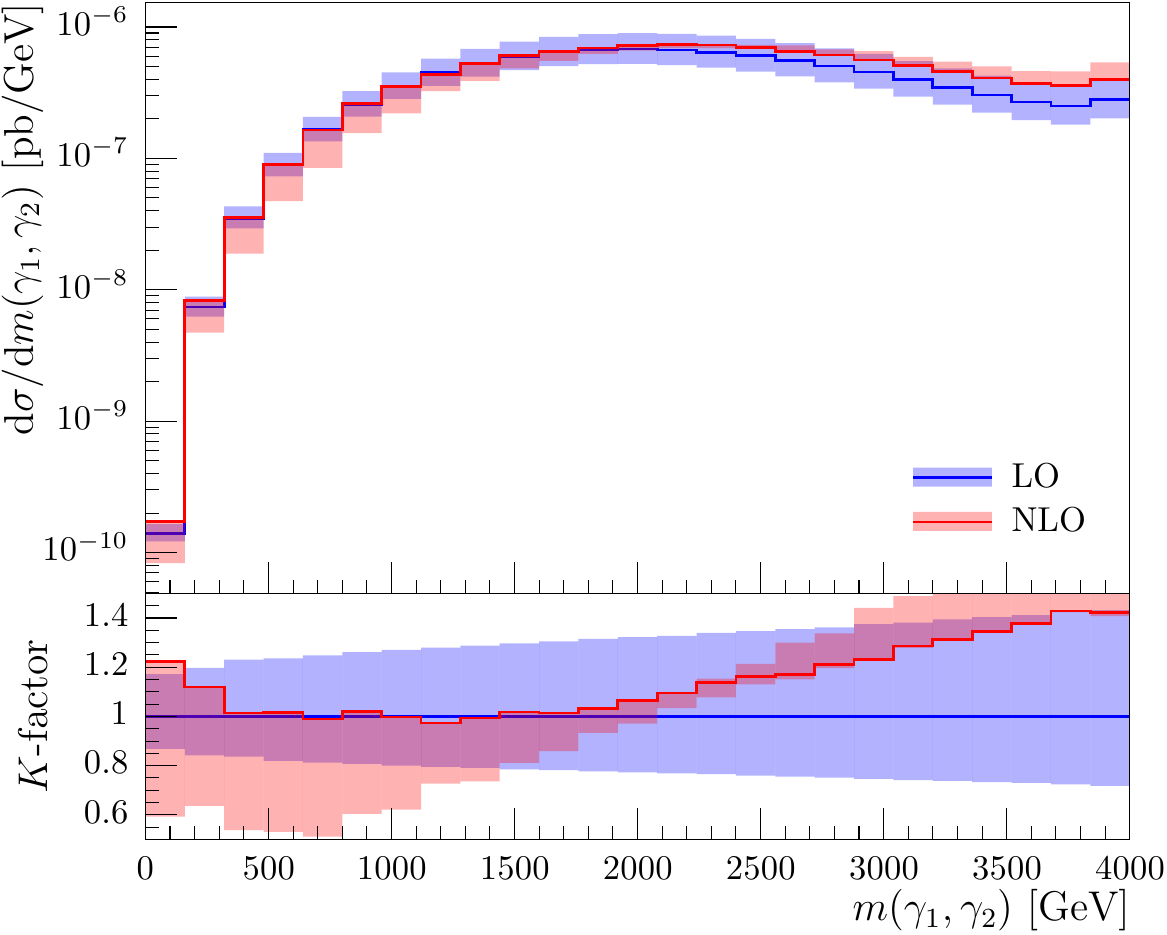}
	\caption{NLO corrections to the invariant di-photon mass distribution stemming from the graviton decay including a scale variation by a factor two around the central scale
	in proton-proton collisions at $\sqrt{s}=8\,\mathrm{TeV}$
	}
	\label{fig:diphotonmass}
\end{figure}

The NLO QCD corrections to the production of a photon pair through graviton exchange in association with a jet
has been calculated with \GoSam within the ADD model \cite{ArkaniHamed:1998rs,Antoniadis:1998ig},
which assumes flat, large extra dimensions.
The model file was generated by
FeynRules \cite{Christensen:2008py} and imported into \GoSam using the UFO file
format \cite{Degrande:2011ua}.  Four to six extra dimensions were assumed and 
an ADD scale $M_S=4\,\mathrm{TeV}$ was chosen. According to this scale, the invariant mass
of the photon pair was restricted to $140\,\mathrm{GeV} \leq m_{\gamma\gamma} \leq 3.99\,\mathrm{TeV}$.
A variation of this upper cut-off showed that the results are not considerably affected by the unknown UV completion of the ADD model.

Since rank-5 boxes and rank-4 triangles appear, the higher rank extensions of \GoSam and \prgname{Golem95C} (cf.
Sec.~\ref{sec:newFeatures}) were used. Additionally, the sum over all discrete Kaluza-Klein mode propagators was approximated by a density function,
which required a new \texttt{custompropagator} extension in \GoSam, and a Lorentz structure where
on-shell conditions were assumed.
All of the needed features are available in the \prgname{Golem95 1.3} and \GoSamTwo release.

For this process, large enhancements are expected in the tail of the diphoton mass spectrum which is shown in Fig.~\ref{fig:diphotonmass}.
In particular, we observe a non-constant K-factor for this observable, which existing experimental searches have not considered.

More details and results can be found in \cite{Greiner:2013gca}, including the case of five and six extra dimensions.

\section{Conclusion}
\GoSamTwo provides many new features, leading to faster and more compact code,
increased usability and a broader range of possible applications.

In particular, \GoSam supports now higher-rank integrals, where the tensor rank exceeds the
number of propagators. This is needed in effective field theories and in calculations with spin-2 particles.
Furthermore, complex masses and couplings, and various electroweak schemes are supported.

A new installation script simplifies the installation of \GoSam and related libraries and tools.

The overall performance was greatly increased by introducing 
a new integrand-level reduction method, implemented in the \prgname{Ninja} library, by
advanced techniques of combining and grouping Feynman diagrams and by using the new \prgname{FORM~4} features for code optimization.

Many of the new features were already successfully used and tested in various phenomenological studies
\cite{Greiner:2012im,vanDeurzen:2013rv,Gehrmann:2013aga,Luisoni:2013cuh,Hoeche:2013mua,Cullen:2013saa,vanDeurzen:2013xla,Gehrmann:2013bga, Dolan:2013rja,Heinrich:2013qaa, Cullen:2012eh, Greiner:2013gca}.

The implementation of the new standardized interface to Monte Carlo programs (BLHA2) allows to use parton showered events
at NLO level and to use different Monte Carlo generators for comparisons.

\section*{Acknowledgements}
The author thanks all the current and former members of the \GoSam collaboration
for the joint effort in developing \GoSamTwo.
Furthermore, he is grateful to the members of the \prgname{Herwig++} collaboration
for the common work on combining both programs via the BLHA2 interface, and Joscha
Reichel for his contribution to the diphoton plus jet through graviton exchange
calculation.

\section*{References}
{
 \makeatletter
 \def\url@#1{\href{http://arxiv.org/abs/#1}{#1}}
 \makeatother 
 \bibliography{gosamrefs}
}

\end{document}